\documentclass[12pt]{article}
\usepackage{amsmath,amssymb,amsfonts}

\newtheorem{theorem}{Theorem}
\newtheorem{prop}[theorem]{Proposition}

\newtheorem{coro}[theorem]{Corollary}
\newtheorem{exe}[theorem]{Exercise}
\newtheorem{exa}[theorem]{Example}

\newtheorem{remark}[theorem]{Remark}
\newenvironment{rem}{\begin{remark} \rm}{\end{remark}}

\newcommand{\tr}{\mathrm{tr}\,}

\newcommand{\CM}{{\cal M}}
\newcommand{\CQ}{{\cal Q}}
\newcommand{\CP}{{\cal P}}

\newcommand{\rref}[1]{(\ref{#1})} 

\newcommand{\del}{{\partial}}

\def\dpt#1#2{{\frac{\partial #1}{\partial #2}}}
\def\endpf{\begin{flushright}$QED$\end{flushright}}

\def\wrt{with respect to}
\def\pd#1{\frac{\partial}{\partial#1}}

\begin{document}
\begin{center}
{\Large \bf On the geometric origin of the bi-Hamiltonian\\ \smallskip
structure of the Calogero-Moser system}
\end{center}
\vspace{0.5truecm}
\begin{center}
C. Bartocci${}^1$, G. Falqui${}^2$, I. Mencattini${}^3$, G. Ortenzi${}^{2}$,
M. Pedroni${}^4$
\end{center}
\vspace{0.1truecm}
\par
\medskip\noindent
{\footnotesize
${}^1$ Dipartimento di Matematica, Universit\`a di Genova,
Via Dodecaneso 35, I-16146 Genova, Italy
\\
${}^2$ Dipartimento di Matematica e Applicazioni,
Universit\`a di Milano-Bicocca,
Via Roberto Cozzi 53, I-20125 Milano, Italy
\\
${}^3$ SISSA, via Beirut 2/4,
I--34014 Trieste, Italy
\\
${}^4$ Dipartimento di Ingegneria dell'Informazione e Metodi Matematici,
Universit\`a di Ber\-ga\-mo,
Viale Marconi 5, I-24044 Dalmine (BG), Italy
\par 
\medskip\noindent
E-mail addresses:\\
bartocci@dima.unige.it, gregorio.falqui@unimib.it,
igorre@sissa.it,\\
giovanni.ortenzi@unimib.it, marco.pedroni@unibg.it}
\vspace{0.1truecm}

\begin{abstract}
We show that the bi-Hamiltonian structure of the rational $n$-particle (attractive)
Calogero-Moser system can be obtained by means of a double projection from a very simple Poisson pair on the cotangent bundle of $\mathfrak{gl}(n,\mathbb{R})$. The relation with the Lax formalism is also discussed.
\end{abstract}

\section{Introduction}

In 1971 Francesco Calogero \cite{Calogero1971} solved the quantum system consisting of $n$ unit-mass particles moving on the line and pairwise interacting via a (repulsive) potential that is proportional to the inverse of the squared distance. (The case $n=3$ was treated earlier \cite{Calogero1969} by Calogero himself). The integrability of the classical counterpart was conjectured in \cite{Calogero1971} and proved by Moser in \cite{Moser}. Later, this system was showed to be superintegrable \cite{rauch}. It is also worthwhile to mention that the classical 3-particle case appeared in the works of Jacobi \cite{Jacobi}. More information on the (quantum and classical) Calogero-Moser system can be found in \cite{Calogero2008}. Recently, this system gained an important role in pure mathematics too. We just cite its relations with quiver varieties \cite{Ginzburg} and double affine Hecke algebras \cite{EtingofGinzburg}, referring to \cite{Etingof} for a more complete list.

Although a lot of papers were devoted to the many facets of the Calogero-Moser system, only a few results concerning its bi-Hamiltonian formulation were found. In \cite{MagMar} (see also \cite{pondi2}) a bi-Hamiltonian structure was constructed with the help of the Lax representation of the system. A $(2n-1)$-dimensional family of compatible Poisson tensors---apparently unrelated with the above mentioned Poisson pair---was found in \cite{goneranutku}, in the context of superintegrable systems. 

In this paper we explain where the bi-Hamiltonian structure of 
\cite{MagMar} comes from. The spirit is very close to that of the fundamental paper \cite{KKS}, where the Calogero-Moser system is shown to be the Marsden-Weinstein reduction of a trivial system on the cotangent bundle of $\mathfrak{su}(n)$. In the same vein, we show that the bi-Hamiltonian structure can be obtained---by means of two projections---from a Poisson pair belonging to a wide class of bi-Hamiltonian structures on cotangent bundles. Such class is recalled in Section 2, while in Section 3 the particular example related to the Poisson brackets of the (attractive) Calogero-Moser system is considered. In Section 4 a first reduction is performed, corresponding to the action given by the simultaneous conjugation. A second projection, leading to the phase space of the Calogero-Moser system, is described in Section 5. Finally, Section 6 is devoted to the example of the 2-particle system (trivial from the physical point of view, but not from the mathematical one) and Section 7 to some final remarks.

\par\smallskip\noindent
{\bf Acknowledgments.}
The authors are grateful to Franco Magri, Andriy Panasyuk, and Volodya Rubtsov for fruitful discussions, and to Victor Ginzburg for some remarks on the archive version of this paper. Moreover, we would like to thank the anonymous referee for useful suggestions, in particular for Remark \ref{rem:ref}.
This work has been partially supported by the European Community through the FP6 Marie Curie RTN {\em ENIGMA} 
(Contract number MRTN-CT-2004-5652) and by the ESF through the research programme MISGAM. M.P. would like to thank for the hospitality the Department  {\em Matematica e Applicazioni\/} of the Milano-Bicocca University, the Department of Mathematics of the Genova University, and the Department of Mathematics of the University of North Carolina at Chapell Hill, where part of this work was done.

\section{Bi-Hamiltonian structures on cotangent bundles}
\label{imm} 

In this section we recall from \cite{Turiel} (see also \cite{IMM}) that a torsionless (1,1) tensor field on a smooth manifold $\CQ$ gives rise to a (second) Poisson structure on the cotangent space $T^*\CQ$, compatible with the canonical one. More information on the geometry of bi-Hamiltonian manifolds can be found, e.g., in \cite{pondi2}.

Let $L:T\CQ\to T\CQ$ be a type $(1,1)$ tensor field on $\CQ$, whose Nijenhuis
torsion vanishes. This means that
\begin{equation}
\label{nijtor}
T(L) (X,Y) := [LX,LY] - L([LX,Y]+ [X,LY] - L[X,Y])=0
\end{equation}
for all pairs of vector fields $X,Y$ on $\CQ$. 
Let $\theta$ be the Liouville
$1$-form on $T^*\CQ$ and $\omega= d\theta$ the standard symplectic
$2$-form on $T^\ast \CQ$, whose associated Poisson tensor will be denoted with $P_0$.
One can deform the
Liouville $1$-form to a $1$-form
$\theta_L$:
$$\langle \theta_L, Z\rangle_\alpha= \langle \alpha, L(\pi_\ast
Z)\rangle_{\pi(\alpha)} \,,$$
for any vector field $Z$ on $T^\ast \CQ$ and for any $1$-form $\alpha$ on
$\CQ$, where $\pi: T^\ast \CQ \to \CQ$ is the canonical
projection. If we choose local coordinates $(x^1,\dots,x^n)$ on $\CQ$ and consider
the corresponding symplectic coordinates $(x^1,\dots,x^n,y_1,\dots,y_n)$ on
$T^\ast \CQ$, we get the local expression $\theta_L = L^i_j y_i dx^j$.
Now, it is well-known that the canonical Poisson bracket is defined by
$$
\{F,G\}_0 = \omega(X_F,X_G)\qquad\qquad F,G\in
C^\infty(T^\ast \CQ)\ ,
$$
where $X_F$, $X_G$ are the Hamiltonian vector fields associated to
$F$, $G$ \wrt\ the symplectic form $\omega$. A second composition law on 
$C^\infty(T^\ast \CQ)$ is given by 
\begin{equation}
\label{secondpb}
\{F,G\}_1 = \omega_L(X_F,X_G)\ ,
\end{equation}
where $\omega_L:=d\theta_L$. It is easily seen that 
$$
\{x^i,x^j\}_1 =0\ ,\quad\{y_i,x^j\}_1 =L_i^j\ ,\quad
\{y_i,y_j\}_1 =\left(\dpt{L_j^k}{x^i}-\dpt{L_i^k}{x^j}\right)y_k\ .
$$
Moreover, the vanishing of the torsion of $L$ entails that \rref{secondpb} is a Poisson bracket too, and that it is compatible with 
$\{\cdot,\cdot\}_0$. Thus we have a bi-Hamiltonian structure on $T^*\CQ$. 

\begin{rem}
Since $P_0$ is invertible, one can introduce the so-called recursion operator 
$N:=P_1 P_0^{-1}$, whose Nijenhius torsion also vanishes. It turns out to be the 
{\it complete lift} of $L$ (see, e.g., \cite{YanoIshihara}), and it is 
uniquely determined by the condition
\begin{equation*}
\label{defdiN}
d\theta_L (X,Y) = \omega(NX,Y)
\end{equation*}
for all vector fields $X$, $Y$ on $T^\ast \CQ$. An easy computation shows that
\begin{align*}
N \left(\pd{x_k}\right) &= L^i_k \pd{x_i} -
y_l\left(\frac{\partial L^l_i}{\partial x_k} - \frac{\partial
L^l_k}{\partial x_i}\right) \pd{y_i}\\
N \left(\pd{y_k}\right) &= L^k_i \pd{y_i}\ .
\end{align*}
As pointed out in \cite{IMM} (see also \cite{BFP,FP} and the references cited therein), the geometry of such bi-Hamiltonian manifolds---often called $\omega N$-manifolds---can be successfully exploited to characterize the Hamiltonian system that are separable in canonical coordinates in which $N$ is diagonal.
\end{rem}

We conclude this section by recalling that the functions 
\begin{equation}
H_k:=\frac1{k}\mathrm{tr}L^k=\frac1{2k}\mathrm{tr}N^k
\end{equation}
form a bi-Hamiltonian hierarchy on $T^*\CQ$, that is,
$P_1 dH_k=P_0 dH_{k+1}$ for all $k\ge 1$. This follows from $N^* dH_k=dH_{k+1}$, where $N^*$ is the transpose of $N$, and is well-known to imply the involutivity (\wrt\ both Poisson brackets) of the $H_k$.

\section{A Bi-Hamiltonian structure on $T^*\mathfrak{gl}(n)$} 
\label{sect3}

In this section we consider a particular case of the general construction described in the previous section. The manifold $\CQ$ is the set $\mathfrak{gl}(n)$ of real $n\times n$ matrices, and the $(1,1)$ torsionless tensor field is defined as
\begin{equation}
L_A:V\mapsto AV\ ,
\end{equation}
where $A\in \mathfrak{gl}(n)$ and $V\in T_A\mathfrak{gl}(n)\simeq\mathfrak{gl}(n)$. 
It is known that the torsion of $L$ vanishes (and one can easily check it by writing \rref{nijtor} for constant vector fields). The cotangent bundle $T^*\mathfrak{gl}(n)\simeq\mathfrak{gl}(n)\times \mathfrak{gl}(n)^*$ can be identified with $\mathfrak{gl}(n)\times \mathfrak{gl}(n)$ by means of the pairing given by the trace of the product. Thus on $\mathfrak{gl}(n)\times \mathfrak{gl}(n)$ we have a bi-Hamiltonian structure, whose first Poisson bracket is associated with the canonical symplectic form
$\omega_0=\tr (dB\wedge dA)$, where $(A,B)\in\mathfrak{gl}(n)\times \mathfrak{gl}(n)$. In order to determine the second Poisson tensor, we have to consider the 1-form $\theta_L=\tr(BA\,dA)$ and to compute
$$
\omega_L=d\theta_L=\tr(dB\wedge A\,dA+B\,dA\wedge dA)\ .
$$
Let $F_1,F_2$ be two real functions on $\mathfrak{gl}(n)\times \mathfrak{gl}(n)$, and let $(\xi_1,\eta_1)$ and $(\xi_2,\eta_2)$ be their differentials. Since the canonical Poisson tensor $P_0$ acts on a covector $(\xi,\eta)$ as
\begin{equation}
\label{poiten0}
P_0:\left(\begin{matrix} \xi\\ \eta\end{matrix}\right)
\mapsto \left(\begin{matrix}
           0 & I\\
          -I & 0
          \end{matrix}\right)
          \left(\begin{matrix} \xi\\ \eta\end{matrix}\right)
          =\left(\begin{matrix} \eta \\ -\xi\end{matrix}\right)\ ,
\end{equation}
the corresponding Hamiltonian vector fields (with respect to $\omega$) are $X_{F_i}=(\eta_i,-\xi_i)$, for $i=1,2$. Therefore
\begin{equation}
\label{5bis}
\{F_1,F_2\}_1=\omega_L(X_{F_1},X_{F_2})=\tr\left(A(\eta_1\xi_2
-\eta_2\xi_1)+B[\eta_1,\eta_2]\right)\ ,
\end{equation}
so that the second Poisson tensor is
\begin{equation}
\label{poiten1}
P_1:\left(\begin{matrix} \xi\\ \eta\end{matrix}\right)
\mapsto \left(\begin{matrix}
           0 & A\cdot\\
          -\cdot A & [B,\cdot]
          \end{matrix}\right)
          \left(\begin{matrix} \xi\\ \eta\end{matrix}\right)
          =\left(\begin{matrix} A\eta \\ -\xi A+[B,\eta]\end{matrix}\right)\ ,
\end{equation}
and the recursion operator $N=P_1 {P_0}^{-1}$ and its transpose are given by
$$
N=\left(\begin{matrix}
           A\cdot & 0\\
           [B,\cdot] & \cdot A
          \end{matrix}\right)\ ,
\qquad
N^*=\left(\begin{matrix}
           \cdot A & [\cdot,B]\\
            0 &  A \cdot
          \end{matrix}\right)\ .
$$
From the expression of $N^*$ it is evident that the functions $H_k=\frac1{k}\tr A^k$, for $k\ge 1$,
form a bi-Hamiltonian hierarchy. It can be easily checked that this hierarchy coincide with the one mentioned at the end of Section 2, namely, that $\tr A^k=\tr L^k$. The corresponding vector fields $X_k:=-P_0 dH_k$ are given by $(X_k)_{(A,B)}=(0,A^{k-1})$.
\begin{rem}
\label{remgroup}
One can check that if $T^*GL(n,\mathbb{R})$ is seen as an open subset of $\mathfrak{gl}(n)\times \mathfrak{gl}(n)$ by means of left translations, then the canonical symplectic structure of $T^*GL(n,\mathbb{R})$ 
takes the form (\ref{poiten1}).
\end{rem}

\begin{rem}
\label{rem:ref}
Another interesting interpretation of $P_1$ is as follows. Let us consider the Cartesian product 
$\mathfrak{gl}(n)\times \mathfrak{gl}(n)$ with the Lie bracket
\begin{equation}
 \label{semidir}
\left[(A_1,B_1),(A_2,B_2)\right]=\left(B_1 A_2-B_2 A_1,[B_1,B_2]\right)\ ,
\end{equation}
i.e., the semidirect product of $\mathfrak{gl}(n)$ with the abelian structure and 
$\mathfrak{gl}(n)$ with its usual Lie bracket. Then $P_1$ is the Lie-Poisson structure associated with 
(\ref{semidir}) on the dual space 
$\left(\mathfrak{gl}(n)\times \mathfrak{gl}(n)\right)^*\simeq
\mathfrak{gl}(n)\times \mathfrak{gl}(n)$, where the identification is given by 
$$
(A,B)\mapsto \tr(A\cdot)+\tr(B\cdot)\ .
$$
Moreover, in the terminology of, e.g., \cite{AK}, $P_0$ is the ``frozen Lie-Poisson structure'' at the point $(I,0)$.
\end{rem}

\section{The first projection} 
\label{sect4}

The main aim of this paper is to show that the bi-Hamiltonian structure of the Calogero-Moser system is a reduction of the one presented in the previous section. The first step is to notice that the Poisson pair $(P_0,P_1)$
is invariant with respect to the action of $G=GL(n,\mathbb{R})$ on $\mathfrak{gl}(n)\times \mathfrak{gl}(n)$ given by the simultaneous conjugation:
$$
\left(g,(A,B)\right)\mapsto \left(gAg^{-1},gBg^{-1}\right)\ .
$$
This is obvious for the canonical Poisson tensor $P_0$, since this action is the lifting to the cotangent bundle of the action $(g,A)\mapsto gAg^{-1}$ on $\mathfrak{gl}(n)$. Since the latter leaves invariant the tensor field $L$, the Poisson tensor $P_1$ is invariant too. We would like to obtain a nice quotient, so we consider $G$ acting on the open subset $\CM\subset \mathfrak{gl}(n)\times \mathfrak{gl}(n)$ formed by the pairs $(A,B)$ such that:
\begin{itemize} 
 \item $A$ and $B$ have real distinct eigenvalues;
 \item if $\{v_i\}_{i=1,\dots,n}$ is an eigenvector basis of $B$, then 
$Av_i\notin \langle v_1,\dots,\hat{v_j},\dots,v_n\rangle$ for all $j\ne i$.
As usual, $\langle \dots \rangle$ denotes the linear span and $\hat{v_j}$ means that ${v_j}$
is not included in the list;
 \item the same condition as before with $A$ and $B$ exchanged.
\end{itemize}
It is clear that $\CM$ is invariant under the action of $G$. Moreover, the description of the quotient $\CM/G$ is made easy by the existence of a subset $\CP\subset \CM$ intersecting every orbit in one point. It is given by $\CP=\bigcup_{\epsilon_i\in\{+,-\}}\CP_{(\epsilon_1,\dots,\epsilon_{n-1})}$, where
$\CP_{(\epsilon_1,\dots,\epsilon_{n-1})}$ is the set of pairs $(A,B)\in \CM$ such that $B$ is diagonal with 
$B_{ii}<B_{jj}$ if $i<j$, and
$A_{i+1,i}>0$, $A_{i,i+1}=\epsilon_i A_{i+1,i}$ for all $i=1,\dots,n-1$.
For example, if $n=2$ we have $\CP=\CP_+\cup \CP_-$, where the elements of $\CP_{+}$ are those in $\CM$ of the form
$$
\left(\begin{pmatrix} A_{11} & A_{21}\\ A_{21} & A_{22}\end{pmatrix},
\begin{pmatrix} B_{11} & 0\\ 0 & B_{22}\end{pmatrix}\right)\ ,
$$ 
with $A_{21}>0$ and $B_{11}<B_{22}$, while the elements of $\CP_{-}$ are those in $\CM$ of the form
$$
\left(\begin{pmatrix} A_{11} & -A_{21}\\ A_{21} & A_{22}\end{pmatrix},
\begin{pmatrix} B_{11} & 0\\ 0 & B_{22}\end{pmatrix}\right)\ ,
$$ 
again with $A_{21}>0$ and $B_{11}<B_{22}$.
\begin{prop}
\label{decomprop}
Every orbit of $G$ in $\CM$ intersects $\CP$ in just one point. Moreover, for all $(A,B)\in \CP$ the tangent space 
$T_{(A,B)}\CM$ is the direct sum of $T_{(A,B)}\CP$ and the tangent space to the orbit.
\end{prop}
{\bf Proof.} Given an orbit of $G$ in $\CM$, it is clearly possible to find a point $(A,B)$ on such orbit with $B$ diagonal and 
\begin{equation}
\label{eigB}
B_{ii}<B_{jj}\qquad \mbox{for all $i<j$.}
\end{equation}
Then, again by the definition of $\CM$, one has that $A_{ij}\ne 0$ if $i\ne j$. Because of \rref{eigB} we can still act on $(A,B)$ only by an invertible diagonal matrix $g=\mbox{diag}(d_1,\dots,d_n)$. We have to show that one can choose the $d_i$ in such a way that $(gAg^{-1},gBg^{-1})=(gAg^{-1},B)\in\CP_{(\epsilon_1,\dots,\epsilon_{n-1})}$ for some $\epsilon_i=\pm 1$. Since $(gAg^{-1})_{ij}=d_i A_{ij} {d_j}^{-1}$, this means that
$$
d_{i+1} A_{i+1,i} {d_i}^{-1}>0\ ,\qquad 
d_{i} A_{i,i+1} {d_{i+1}}^{-1}=\epsilon_i d_{i+1} A_{i+1,i} {d_i}^{-1}\ .
$$
Therefore $\epsilon_i$ is determined by the sign of $A_{i,i+1}/A_{i+1,i}$, and 
$$
\frac{d_i}{d_{i+1}}=\pm \sqrt{\epsilon_i\frac{A_{i+1,i}}{A_{i,i+1}}}\ ,
$$
where the $\pm$ has to be chosen in such a way that $d_{i+1} A_{i+1,i} {d_i}^{-1}>0$. In this way we have found the matrix $g$ up to a multiple, and the first part of the claim is proved.

Let us fix now a point $(A,B)\in\CP_{(\epsilon_1,\dots,\epsilon_{n-1})}$ and a tangent vector $(V,W)\in T_{(A,B)}\CM$. We have to show that $(V,W)$ can be uniquely decomposed as
\begin{equation}
\label{decomp}
(V,W)=(\dot A,\dot B)+([A,\xi],[B,\xi])\ ,
\end{equation}
where $(\dot A,\dot B)\in T_{(A,B)}\CP_{(\epsilon_1,\dots,\epsilon_{n-1})}$ and $\xi\in\mathfrak{gl}(n,\mathbb{R})$. Since $\dot B$ is diagonal, we immediately have that the off-diagonal entries of $\xi$ are given by $\xi_{ij}=W_{ij}/(B_{ii}-B_{jj})$. Then we have to impose the conditions 
${\dot A}_{i,i+1}=\epsilon_i {\dot A}_{i+1,i}$,
getting the following equations,
$$
\sum_{j=1}^n\left(\xi_{ij}{A}_{j,i+1}-A_{ij}{\xi}_{j,i+1}\right)
=\epsilon_i
\sum_{j=1}^n\left(\xi_{i+1,j}{A}_{ji}-A_{i+1,j}{\xi}_{ji}\right)
-{V}_{i,i+1}+\epsilon_i V_{i+1,i}\ ,
$$
for all $i=1,\dots,n-1$.
Since ${A}_{i,i+1}=\epsilon_i {A}_{i+1,i}$, we obtain
\begin{eqnarray*}
2\epsilon_i {A}_{i+1,i}(\xi_{ii}-\xi_{i+1,i+1})=
&-&\sum_{j\ne i}\left(\xi_{ij}{A}_{j,i+1}+\epsilon_i\xi_{ji}{A}_{i+1,j}\right)\\
&+&\sum_{j\ne i+1}\left(\xi_{j,i+1}{A}_{ij}+\epsilon_i\xi_{i+1,j}{A}_{ji}\right)
-{V}_{i,i+1}+\epsilon_i V_{i+1,i}\ ,
\end{eqnarray*}
for all $i=1,\dots,n-1$,
namely, $(n-1)$ equations for the variables $\xi_{11},\dots,
\xi_{nn}$. Thanks to the fact that ${A}_{i+1,i}>0$, they can be solved, and the solution is unique up to a (common) additive constant. This shows the uniqueness of the vector $([A,\xi],[B,\xi])$, tangent to the orbit, and of the decomposition \rref{decomp}.
\endpf

\begin{rem}
It is clear from the previous proof that one can also identify $\CM/G$ with the submanifold $\CP'\subset\CM$, whose definition is the same of $\CP$, but with the matrices $A$ and $B$ exchanged. 
\end{rem}

\begin{rem}
The quotient space defined by simultaneous conjugation on $k$-tuples of matrices has been the subject of important investigations by Artin, Procesi, Razmyslov, and others (see, e.g., \cite{lebruyn}). For our purposes, it is convenient to restrict to the open subset $\CM$, and in this case an explicit description of the quotient is possible in terms of the transversal submanifold $\CP$.
\end{rem}

Next we consider the vector fields $X_k$ of the bi-Hamiltonian hierarchy on $\CM$, that is, $\left(X_k\right)_{(A,B)}=
\left(0,A^{k-1}\right)$. Since their Hamiltonians $H_k=\frac{1}{k}\tr A^k$ and 
the bi-Hamiltonian structure are invariant with respect to the action of $G$, the $X_k$ can be projected on $\CM/G$. Their projections are the vector fields associated with the Hamiltonians $H_k$ (seen as functions on the quotient) and the reduced bi-Hamiltonian structure. We can exploit the identification between $\CM/G$ and the submanifold $\CP\subset\CM$ in order to explicitly write the projected vector fields. 
Indeed, we have just seen that we can uniquely find $(\del_k A,
\del_k B)\in T_{(A,B)}\CP$ and $([A,\xi_k],[B,\xi_k])$, tangent to the orbit passing through $(A,B)\in\CP$, such that 
$$
(0,A^{k-1})=(\del_k A,\del_k B)+([A,\xi_k],[B,\xi_k])\ .
$$
This shows that
\begin{equation}
\label{LaxP}
\del_k A=[\xi_k,A]\ ,\qquad \del_k B=[\xi_k,B]+A^{k-1}\ ,
\end{equation}
i.e., the projected flows possess a Lax representation. 
In the next section we will perform a second reduction and we will show that the flows \rref{LaxP} give rise to the (attractive) Calogero-Moser flows.

\begin{rem}
The deduction of the Lax equations (\ref{LaxP}) is well-known (see, e.g., \cite{Perelomov}). Notice however that such equations describe flows on the $(n^2+1)$-dimensional manifold $\CP\simeq\CM/G$ and so they are an extension of the usual Lax representation of the Calogero-Moser system.
\end{rem}

We close this section with an interesting description of the 
bi-Hamiltonian structure on the quotient $\CM/G$ (see \cite{Etingof}, where only the first Poisson structure is considered). 
Let $F_1=\tr(a_1\cdots a_r)$ and 
$F_2=\tr(b_1\cdots b_s)$, where $a_i$ and $b_j$ are either $A$ or $B$. Then 
$$
dF_1=\left(\sum_{i:a_i=A}a_{i+1}\cdots a_r a_1\cdots a_{i-1},
\sum_{j:a_j=B}a_{j+1}\cdots a_r a_1\cdots a_{j-1}
\right)
$$
and therefore we have the so-called necklace bracket formula
\begin{equation}
\begin{aligned}
\label{necklace0}
\{F_1,F_2\}_0&=\sum_{(i,j):a_i=B,b_j=A}\tr\left(a_{i+1}\cdots a_r a_1\cdots a_{i-1}
b_{j+1}\cdots b_s b_1\cdots b_{j-1}\right)\\
&-\sum_{(i,j):a_i=A,b_j=B}\tr\left(b_{j+1}\cdots b_s b_1\cdots b_{j-1}a_{i+1}\cdots a_r a_1\cdots a_{i-1}\right)\ .
\end{aligned}
\end{equation}
As far as the second Poisson bracket is concerned, we have from (\ref{5bis}) that
\begin{equation}
\begin{aligned}
\label{necklace1}
\{F_1,F_2\}_1&=\sum_{(i,j):a_i=B,b_j=A}\tr\left(A a_{i+1}\cdots a_r a_1\cdots a_{i-1}
b_{j+1}\cdots b_s b_1\cdots b_{j-1}\right)\\
&-\sum_{(i,j):a_i=A,b_j=B}\tr\left(A b_{j+1}\cdots b_s b_1\cdots b_{j-1}a_{i+1}\cdots a_r a_1\cdots a_{i-1}\right)\\
&+\sum_{(i,j):a_i=B,b_j=B}\tr\left(B[a_{i+1}\cdots a_r a_1\cdots a_{i-1},
b_{j+1}\cdots b_s b_1\cdots b_{j-1}]\right)\\
&=\sum_{(i,j):a_i=B,b_j=A}\tr\left(a_i a_{i+1}\cdots a_r a_1\cdots a_{i-1}
b_{j+1}\cdots b_s b_1\cdots b_{j-1}\right)\\
&-\sum_{(i,j):a_i=A,b_j=B}\tr\left(b_{j+1}\cdots b_s b_1\cdots b_{j-1}a_{i+1}\cdots a_r a_1\cdots a_{i-1}a_i\right)\\
&+\sum_{(i,j):a_i=B,b_j=B}\tr\left(a_i a_{i+1}\cdots a_r a_1\cdots a_{i-1}
b_{j+1}\cdots b_s b_1\cdots b_{j-1}\right)\\
&-\sum_{(i,j):a_i=B,b_j=B}\tr\left(b_j b_{j+1}\cdots b_s b_1\cdots b_{j-1}a_{i+1}\cdots a_r a_1\cdots a_{i-1}\right)\ .
\end{aligned}
\end{equation}

Now let us pass from the $(n^2+1)$-dimensional quotient $\CM/G\simeq\CP$ to the phase space of the Calogero-Moser system.

\section{The second projection}

In this section we will perform a second reduction of the bi-Hamiltonian structure on $\CM/G\simeq \CP$. The starting point is the observation that the invariant functions
$$
I_k(A,B)=\frac1{k}\tr A^k=H_k(A,B)\ ,\quad J_k(A,B)=\tr(A^{k-1}B)\ ,\quad \mbox{for $k=1,\dots,n$,}
$$
form a Poisson subalgebra with respect to both Poisson brackets. Indeed, by direct computation or using the necklace bracket formulas (\ref{necklace0}-\ref{necklace1}), one finds that
\begin{equation}
\label{magmar}
\begin{array}{c}
\{I_k,I_l\}_0=0\ ,\quad \{J_l,I_k\}_0=(k+l-2)I_{k+l-2}\ ,\quad \{J_k,J_l\}_0=(l-k)J_{k+l-2}\ ,\\
\{I_k,I_l\}_1=0\ ,\quad \{J_l,I_k\}_1=(k+l-1)I_{k+l-1}\ ,\quad \{J_k,J_l\}_1=(l-k)J_{k+l-1}\ ,
\end{array}
\end{equation}
with the exception that $\{J_1,I_1\}_0=n$. In any case, the Cayley-Hamilton theorem implies that all the right-hand sides of (\ref{magmar}) can be written in terms of $I_1,\dots,I_n,J_1,\dots,J_n$. Thus both Poisson brackets can be further projected on the quotient space defined by the map 
$\pi:\CM/G\simeq \CP\to \mathbb{R}^{2n}$ whose components are the functions $I_1,\dots,I_n,J_1,\dots,J_n$. 

\begin{prop}
 The map $\pi$ is a submersion, i.e., its differential is surjective at every point of $\CM/G$. 
\end{prop}
{\bf Proof.} It is convenient to identify $\CM/G$ with $\CP'$ and to consider, among the coordinates on $\CP'$, the diagonal entries $(\lambda_1,\dots,\lambda_n)$ of the (diagonal) matrix $A$ and the diagonal entries $(\mu_1,\dots,\mu_n)$ of $B$. Then 
$$
I_k=\frac1{k}\sum_{l=1}^n {\lambda_l}^k\ ,\qquad J_k=\sum_{l=1}^n \mu_l{\lambda_l}^{k-1}\ ,
$$
which implies that 
$$
\det\left(\begin{matrix} \frac{\partial I}{\partial\lambda} & \frac{\partial J}{\partial\lambda}\\
\frac{\partial I}{\partial\mu} & \frac{\partial J}{\partial\mu}
\end{matrix}\right)\ne 0
$$
since the $\lambda_i$ are distinct. This shows that the differential of $\pi$ is surjective.
\endpf

The previous proposition entails that the image of $\pi$ is an open subset $U\subset{\mathbb{R}}^{2n}$,
which is in 1-1 correspondence with the second quotient space (by its very definition).
Our final step is to prove that the projection on $U$ gives rise to the phase space of the attractive Calogero-Moser system, with its bi-Hamiltonian flows. To do this, we recall once more that the (first) quotient $\CM/G$ can be identified with the submanifold $\CP\subset\CM$ and we restrict to its connected component 
$\CP_{(-,\dots,-)}$, that is, the set of pairs $(A,B)\in \CM$ such that $B$ is diagonal with 
$B_{ii}<B_{jj}$ if $i<j$, and $A_{i+1,i}>0$, $A_{i,i+1}=-A_{i+1,i}$ for all $i=1,\dots,n-1$. Then we
introduce a submanifold $\CQ\subset\CP_{(-,\dots,-)}$ which will be shown to be in 1-1 correspondence with an open subset of the second quotient space. The elements of $\CQ$ are the pairs $(L,\mbox{diag}(x_1,\dots,x_n))\in\CP$ such that 
$x_{i}<x_{j}$ if $i<j$, and $L_{ij}=\frac1{x_{i}-x_{j}}$ if $i\ne j$. 
If we put $L_{ii}=y_i$, we obtain the Lax matrix of the attractive Calogero-Moser system:
$$
L=\left(\begin{matrix} y_1 & \frac1{x_{1}-x_{2}} & \cdots & \frac1{x_{1}-x_{n}}\\
\frac1{x_{2}-x_{1}} & y_2 & \ddots & \vdots\\
\vdots & \ddots & \ddots & \vdots\\
\frac1{x_{n}-x_{1}} & \frac1{x_{n}-x_{2}} & \cdots & y_n
        \end{matrix}\right)\ .
$$
Let us also introduce the submanifold $\CQ'\subset\CP'$ whose elements are the pairs $(\mbox{diag}(\lambda_1,\dots,\lambda_n),L')\in\CP'$ such that 
$\lambda_{i}<\lambda_{j}$ if $i<j$, and $L'_{ij}=\frac1{\lambda_{j}-\lambda_{i}}$ if $i\ne j$. 
In order to identify $\CQ$ with a subset of the second quotient space, we need the following result. It is a restatement of Proposition 2.6 in \cite{Etingof}, but we give its proof for the reader's sake. 
\begin{prop}
\label{propetingo}
If $\rho:\CM\to\CM/G$ is the canonical projection, then 
$\rho(\CQ)$ coincides with $\rho(\CQ')$, and is formed by the orbits of the pairs $(A,B)$ such that the rank of $[B,A]+I$ is 1.
\end{prop}
{\bf Proof.} We notice that 
$$
\CQ=\{(A,B)\in\CP\mid [B,A]=\mu\}\ \mbox{and}\ \CQ'=\{(A,B)\in\CP'\mid [B,A]=\mu\}\ ,
$$ 
where $\mu_{ij}=1-\delta_{ij}$. Thus, the elements $(A,B)$ in the orbits passing through $\CQ$ and $\CQ'$ satisfy the condition $\mbox{rank}([B,A]+I)=1$. Conversely, let us suppose that $(A,B)\in\CM$ and the above condition 
holds. We can also suppose that $B$ has already been diagonalized. Since the rank of $K:=[B,A]+I$ is 1, there exist $a_i,b_i\in\mathbb{R}$, $i=1,\dots,n$, such that $K_{ij}=a_i b_j$. From $[B,A]_{ij}=(B_{ii}-B_{jj})A_{ij}$ we have that $K_{ii}=1$ and therefore $b_i={a_i}^{-1}$. By acting with $\mbox{diag}(a_1,\dots,a_n)$, the entries of the matrix $K$ all become 1 and so $(A,B)$ is mapped into $\CQ$. This shows that the orbits in $\rho(\CQ)$ are precisely those of the pairs $(A,B)$ such that $\mbox{rank}([B,A]+I)=1$. Diagonalizing $A$ instead of $B$, one proves that the same is true for $\rho(\CQ')$.
\endpf

\begin{coro}
The restriction to $\CQ$ of the map $\pi=(I_1,\dots,I_n,J_1,\dots,J_n)$ is injective.
\end{coro}
{\bf Proof.} Since $\pi$ is an invariant map, we can exploit the identification between $\CQ$ and $\CQ'$ given by the previous proposition and show that $\pi$ is injective on $\CQ'$. If $(D',L')\in\CQ'$, with 
$$
D'=\mbox{diag}(\lambda_1,\dots,\lambda_n)\ ,\qquad
L'=\left(\begin{matrix} \mu_1 & \frac1{\lambda_{2}-\lambda_{1}} & \cdots & \frac1{\lambda_{n}-\lambda_{1}}\\
\frac1{\lambda_{1}-\lambda_{2}} & \mu_2 & \ddots & \vdots\\
\vdots & \ddots & \ddots & \vdots\\
\frac1{\lambda_{1}-\lambda_{n}} & \frac1{\lambda_{2}-\lambda_{n}} & \cdots & \mu_n
        \end{matrix}\right)\ ,
$$
then 
$$
I_k(D',L')=\frac1{k}\sum_{l=1}^n {\lambda_l}^k\ ,\qquad J_k(D',L')=\sum_{l=1}^n \mu_l{\lambda_l}^{k-1}\ .
$$
To conclude that $\pi$ is injective, we simply have to recall that $\lambda_i<\lambda_j$ if $i<j$.
\endpf

\begin{rem}
 From the proof of Proposition \ref{propetingo} it is also clear that $$
\rho(\CQ)=\rho(\CQ')=\{\mbox{orbits of 
the pairs $(A,B)$ such that $[B,A]=\mu$}\} .
$$ 
\end{rem}

We have thus shown that an open subset of the quotient defined by the map $\pi$ can be identified with the submanifold $\CQ$, that is, the phase space of the Calogero-Moser system. The 
bi-Hamiltonian structure on $\CQ$ is given by the Poisson brackets \rref{magmar}.

\begin{rem}
Formulas \rref{magmar} appeared in \cite{MagMar} (see also \cite{pondi2}), in the context of the repulsive Calogero-Moser system. In that paper the construction of the bi-Hamiltonian structure on the Calogero-Moser phase space starts from the Lax representation of the system and uses a special class of coordinates defined on regular bi-Hamiltonian manifolds (the 
so-called Darboux-Nijenhuis coordinates, see \cite{MagMar,pondi2,FP}).
On the contrary, here we recover both the Poisson brackets \rref{magmar} and the Lax representation from the Poisson pair (\ref{poiten0}-\ref{poiten1}) 
on $\mathfrak{gl}(n)\times \mathfrak{gl}(n)$. 
\end{rem}

Now we consider the (projected) bi-Hamiltonian hierarchy on $\CP_{(-,\dots,-)}\subset\CM/G$. Since the Hamiltonians are just the functions $H_k=I_k$, this hierarchy further projects on the second quotient space. In particular, it gives rise to bi-Hamiltonian vector fields on $\CQ$, which we will soon see to be those of the attractive Calogero-Moser system. In principle, to write these vector fields we should project the flows \rref{LaxP}, as we did after Proposition \ref{decomprop}. But they are already tangent to $\CQ$, as shown in

\begin{prop}
\label{tangentprop}
Let $(A,B)\in\CQ$ and let $(\del_k A,\del_k B)\in
T_{(A,B)}\CP$ be given by \rref{LaxP}, that is,
\[
\del_k A=[\xi_k,A]\ ,\qquad \del_k B=[\xi_k,B]+A^{k-1}\ .
\]
Then $(\del_k A,\del_k B)\in T_{(A,B)}\CQ$ for all $k\ge 1$.
\end{prop}
{\bf Proof.}
We know that 
$$
\CQ=\{(A,B)\in\CP\mid [B,A]=\mu\}\ ,
$$ 
where $\mu_{ij}=1-\delta_{ij}$. This entails that $(\del_k A,\del_k B)\in
T_{(A,B)}\CQ$ if and only if $[\del_k A,B]+[A,\del_k B]=0$ at the points of $\CQ$. But this is equivalent to the 
assertion that $[\xi_k,\mu]=0$.
Let us prove this fact, introducing a matrix $\xi$ such that 
$\xi_{ij}=(\xi_k)_{ij}=\left(A^{k-1}\right)_{ij}/(x_i-x_j)$ for $i\ne j$, and 
\begin{equation}
\label{xidiag}
\xi_{ii}=-\frac12 \sum_{l\ne i}(\xi_{il}+\xi_{li})\ . 
\end{equation}
Since $B$ is diagonal, we have that 
\begin{equation}
 \label{xi-xik}
\del_k B=[\xi_k,B]+A^{k-1}=[\xi,B]+A^{k-1}\ .
\end{equation} 
At the end, it will turn out that $\xi$ is a possible choice for $\xi_k$ (recall from the proof of Proposition \ref{decomprop} that $\xi_k$ is determined up to a 
multiple of the identity matrix). Now let us show that $[\xi,\mu]=0$. Indeed,
$$
[\xi,\mu]=[\xi,[B,A]]=[A,[B,\xi]]+[B,[\xi,A]]=-[A,\del_k B]+[B,[\xi,A]]\ ,
$$
so that, putting $\del_k B=\mbox{diag}({\dot x}_1,\dots,{\dot x}_n)$, we have
\begin{equation}
 \label{commij}
[\xi,\mu]_{ij}=({\dot x}_i-{\dot x}_j)A_{ij}+(x_i-x_j)[\xi,A]_{ij}\ .
\end{equation}
Thus $[\xi,\mu]_{ii}=0$ for all $i=1,\dots,n$. Moreover, (\ref{xidiag}) implies that, for $i\ne j$,
$$
[\xi,\mu]_{ij}=\frac12 \sum_{l}(\xi_{il}-\xi_{li})+\frac12 \sum_{m}(\xi_{jm}-\xi_{mj})=[\xi,\mu]_{ii}+[\xi,\mu]_{jj}\ .
$$
Therefore we have that $[\xi,\mu]=0$. In order to finish the proof, we have to show that $\xi$ is a possible choice for $\xi_k$, i.e., that 
$[\xi,A]_{i,i+1}=-[\xi,A]_{i+1,i}$ for all $i=1,\dots,n-1$. But from (\ref{commij}), the vanishing of $[\xi,\mu]$, and the fact that $A_{ij}=-A_{ji}$ for $i\ne j$, we have that
$$
0=(x_i-x_j)\left([\xi,A]_{ij}+[\xi,A]_{ji}\right)\ 
$$
and therefore $[\xi,A]_{ij}=-[\xi,A]_{ji}$.
\endpf

As it is well-known, the Calogero-Moser system is given by the second flow $\del_2 A=[\xi_2,A]$, where $(\xi_2)_{ij}=1/(x_i-x_j)^2$ for $i\ne j$ and 
$(\xi_2)_{ii}=-\sum_{j\ne i}(\xi_2)_{ij}$. In general $\xi_k$ is not symmetric for $k>2$.

\begin{rem}
In the repulsive case, corresponding to the connected component $\CP_{(+,\dots,+)}\subset \CM/G$, one can still introduce the submanifold 
$$
\CQ_+=\left\{(L,\mbox{diag}(x_1,\dots,x_n))\mid x_{i}<x_{j}
\mbox{ and } L_{ij}=L_{ji}=\frac1{x_{i}-x_{j}}
\mbox{ if $i<j$}\right\}\ 
$$
and show that it can be identified with (an open subset of) the second quotient space. However, if $n>2$ the flows \rref{LaxP} are not tangent to $\CQ_+$ and therefore they need to be projected on $\CQ_+$, where they assume a more complicated form.
\end{rem}
\begin{rem}
It is well-known \cite{KKS,Etingof,Wilson} that the first Poisson structure $P_0$ can be reduced on the phase space of the Calogero-Moser system by means of the Marsden-Weinstein reduction and that it gives rise to the canonical structure in the coordinates $(x_i,y_j)$. Our reduction consists in a double projection, and has the same effect on $P_0$. 
However, we can reduce also the second Poisson structure $P_1$, on which the Marsden-Weinstein reduction cannot be performed, since it employs the moment map $(A,B)\to [A,B]$ of $P_0$. Notice that this moment map appears in the proof of Proposition \ref{propetingo}.
\end{rem}
\begin{rem}
We have used two projections to obtain the bi-Hamiltonian structure of the Calogero-Moser system from the one on $T^*\mathfrak{gl}(n)$. Of course, such projections can be composed and we could have found directly the Poisson pair on $\CQ$. We decided to study also the first quotient $\CM/G$ because it is more natural and we think that it might be of interest on its own.
\end{rem}

\section{Example: $n=2$}

In this section we consider the 2-particle Calogero-Moser system in order to exemplify our construction. The starting point is the set $\CM$ whose elements are pairs $(A,B)$ of matrices in $\mathfrak{gl}(2,\mathbb{R})$ such that $A$ and $B$ have real distinct eigenvalues and no common eigenvector. The first quotient space $\CM/G$ can be identified with the 5-dimensional manifold $\CP=\CP_+\cup \CP_-$ (see Section \ref{sect4}) or with $\CP'=\CP'_+\cup \CP'_-$, 
where the elements of $\CP'_{+}$ are those in $\CM$ of the form
$$
\left(\begin{pmatrix} A_{11} & 0\\ 0 & A_{22}\end{pmatrix},
\begin{pmatrix} B_{11} & B_{12}\\ B_{12} & B_{22}\end{pmatrix}\right)\ ,
$$ 
with $A_{11}<A_{22}$ and $B_{12}>0$, while the elements of $\CP'_{-}$ are 
those in $\CM$ of the form
$$
\left(\begin{pmatrix} A_{11} & 0\\ 0 & A_{22}\end{pmatrix},
\begin{pmatrix} B_{11} & B_{12}\\ -B_{12} & B_{22}\end{pmatrix}\right)\ ,
$$ 
again with $A_{11}<A_{22}$ and $B_{12}>0$. There are only two independent vector fields $X_1$ and $X_2$ in the  bi-Hamiltonian hierarchy on $\CM$, corresponding to the invariant functions $H_1=\tr A$ and $H_2=\frac1{2}\tr A^2$. 
The equations on $\CM/G\simeq\CP$ are given by (\ref{LaxP}), with $\xi_1=I$ and 
$$
\xi_2=\begin{pmatrix} 0 & \frac{A_{21}}{B_{22}-B_{11}}\\ \frac{A_{21}}{B_{22}-B_{11}} & 0\end{pmatrix}
\ \mbox{on $\CP_-$},\quad
\xi_2=\begin{pmatrix} 0 & -\frac{A_{21}}{B_{22}-B_{11}}\\ \frac{A_{21}}{B_{22}-B_{11}} & 0\end{pmatrix}
\ \mbox{on $\CP_+$}.
$$ 
The second projection $\pi:\CP\to\mathbb{R}^4$ is given by $\pi=(I_1,I_2,J_1,J_2)$, where
$$
I_1=H_1=\tr A,\quad I_2=H_2=\frac1{2}\tr A^2,\quad 
J_1=\tr B,\quad J_2=\tr(AB)\ .
$$
One can check that the image of $\pi$ is the set
$$
U=\left\{(I_1,I_2,J_1,J_2)\in\mathbb{R}^4\mbox{ s.t. } 
4I_2-{I_1}^2>0, J_2-\frac12 I_1J_1\ne 0
\right\}\ .
$$
The restriction of $\pi$ to
$$
\CQ=\left\{\left(
\left(\begin{matrix} y_1 & \frac1{x_{1}-x_{2}}\\ 
\frac1{x_{2}-x_{1}} & y_2 \end{matrix}\right),
\left(\begin{matrix} x_1 & 0\\ 
0 & x_2 \end{matrix}\right)\right)\mbox{ s.t. } 
x_1<x_2, |y_1-y_2|(x_2-x_1)>2
\right\}
$$
is a bijection onto
$$
V=\left\{(I_1,I_2,J_1,J_2)\in\mathbb{R}^4\mbox{ s.t. } 
4I_2-{I_1}^2>0, |J_2-\frac12 I_1J_1|>1
\right\}\ .
$$
The Poisson brackets on $\CQ$ are given by
\begin{equation}
\label{magmar2}
\begin{array}{l}
\{I_1,I_2\}_0=0,\quad \{J_1,I_1\}_0=2,
\quad \{J_1,I_2\}_0=\{J_2,I_1\}_0=I_1,\\
\quad \{J_2,I_2\}_0=2I_2,\quad \{J_1,J_2\}_0=J_{1},\\
\{I_1,I_2\}_1=0,\quad \{J_1,I_1\}_1=I_1,
\quad \{J_1,I_2\}_1=\{J_2,I_1\}_1=2I_2,\\
\quad \{J_2,I_2\}_1=3I_3=3I_1 I_2-\frac12 {I_1}^3,
\quad \{J_1,J_2\}_1=J_{2}.
\end{array}
\end{equation}
In terms of the physical coordinates $(x_1,x_2,y_1,y_2)$, the first bracket is the canonical one, while
\begin{equation}
\label{magmar2xy}
\begin{array}{l}
\{x_1,x_2\}_1=\frac{2x_{12}}{\Delta},\quad 
\{x_1,y_1\}_1=y_1+\frac{(y_1-y_2){x_{12}}^2}{\Delta},\quad 
\quad \{x_2,y_2\}_1=y_2-\frac{(y_1-y_2){x_{12}}^2}{\Delta},
\\
\{y_2,x_1\}_1=\{x_2,y_1\}_1=\frac{(y_1-y_2){x_{12}}^2}{\Delta},\quad 
\{y_1,y_2\}_1=-{x_{12}}^3.
\end{array}
\end{equation}
where $x_{12}=1/(x_1-x_2)$ and $\Delta=4{x_{12}}^2-(y_1-y_2)^2$.
Notice that the term $\Delta$ appearing in the denominators of 
(\ref{magmar2xy}) is the discriminant of the characteristic polynomial of 
$$
\left(\begin{matrix} y_1 & \frac1{x_{1}-x_{2}}\\ 
\frac1{x_{2}-x_{1}} & y_2 \end{matrix}\right)\ .
$$
This means that one cannot reduce the second Poisson bracket on the whole Calogero-Moser phase space, but only on its open subset $\CQ$.

\section{Final remarks}

In this paper we have shown that the Poisson pair of the (rational, attractive) Calogero-Moser system is a reduction of a very natural bi-Hamiltonian structure on $T^*\mathfrak{gl}(n,\mathbb{R})$. A first possible development of this result is to extend this construction to other Calogero-Moser systems, such as the trigonometric one (see also Remark \ref{remgroup}) and those associated to (root systems of) simple Lie algebras \cite{Perelomov}. Secondly, it would be interesting to investigate, from the point of view of bi-Hamiltonian geometry, the problem of duality between Calogero-Moser systems \cite{Ruij,FGNR} and, more generally, between integrable systems. In particular, on $\CM$ there is another bi-Hamiltonian structure, obtained by exchanging $A$ with $B$. (In other words, one can look at $B$ as the ``point'' in $\mathfrak{gl}(n,\mathbb{R})$ and $A$ as the ``covector'' in $T^*_B\mathfrak{gl}(n,\mathbb{R})\simeq\mathfrak{gl}(n,\mathbb{R})$, and consider the (1,1) tensor field $V\mapsto BV$ on $\mathfrak{gl}(n,\mathbb{R})$.) The functions $H_k'=\frac1{k}\tr B^k$, for $k\ge 1$,
form a bi-Hamiltonian hierarchy with respect to the new Poisson pair. Being invariant with respect to the action of $G$, such pair can be projected on the (first) quotient space $\CM/G$, along with its hierarchy. However, they 
cannot be projected on the second quotient space, but one has to introduce the map 
$\pi':\CM/G\to \mathbb{R}^{2n}$ whose components are the functions
$$
I'_k(A,B)=\frac1{k}\tr B^k=H'_k(A,B)\ ,\quad J'_k(A,B)=\tr(AB^{k-1})\ ,\quad \mbox{for $k=1,\dots,n$.}
$$
The projection along $\pi'$ gives rise again to the rational Calogero-Moser system, which is indeed well-known to be dual to itself. From the bi-Hamiltonian viewpoint, it is important to observe that the map $(A,B)\mapsto (-B,A)$ sends the Poisson pair $(P_0,P_1)$ into the new Poisson pair.

\end{document}